# A Cloud Condensation Model on Insoluble Nuclei Validated with Radiosonde and Ceilometer Data: Implications for Dust based Rainfall Forecasting

**Rani Arielly[a], Adva Baratz[b], Ran Aharoni[c] and Ofir Shoshanim[d]**

[a] Institute of Agricultural and Biosystems Engineering, Agricultural Research Organization - Volcani Institute, Rishon LeZion, 7505101, Israel. Email: rania@volcani.agri.gov.il

[b] Department of Analytical Chemistry, Israel Institute for Biological Research, Ness Ziona, 74100, Israel. Email: adva.baratz@gmail.com

[c] Department of Physics, Nuclear Research Center, Negev, Beer-Sheva, 84109, Israel. Email: tenzoran@gmail.com

[d] Department of Environmental Physics, Israel Institute for Biological Research, Ness-Ziona, 74100, Israel. Email: ofirshoshanim@gmail.com

Corresponding author: Rani Arielly (rania@volcani.agri.gov.il)

## Abstract

This study introduces and validates a cloud condensation model on insoluble nuclei using a novel comparative analysis of simultaneous radiosonde and ceilometer data. A transformation of the radiosonde's temperature and relative humidity profiles into a simulated optical backscatter signal is implemented via the model, which includes a critical rate-limiting activation step. By comparing this simulated signal with the ceilometer's measured data, the model can determine the minimum effective size of dust particles required to act as cloud condensation nuclei. This approach has direct implications for improving the parameterization of cloud formation in local weather models, particularly for rainfall forecasting in dust-prone arid and semi-arid regions.

## Glossary

atmospheric boundary layer (ABL): The lowest part of the troposphere, directly influenced by surface forces such as evaporation, transpiration, and local pressure gradients cloud condensation nuclei (CCN): Small particles in the atmosphere (such as dust, salt, or pollutants) that serve as surfaces for water vapor to condense upon, facilitating cloud droplet formation.

Light Detection and Ranging (LiDAR): A remote sensing technology that uses pulsed laser light to measure distances and detect the presence of matter.

# 1    Introduction

Accurate, localized prediction of precipitation remains a significant challenge in operational meteorology, particularly in arid and semi-arid regions where aerosol cloud precipitation interactions are dominated by mineral dust. Operational weather forecasts often struggle with precipitation accuracy when neglecting the influence of aerosols, which can lead to over-forecasting light rain and under-forecasting heavy rain events (Jiang et al., 2017). The concentration, size, and composition of Cloud Condensation Nuclei (CCN) are critical variables that significantly impact cloud microphysical properties, droplet spectra, and ultimately, the efficiency of rain production (Cecchini et al., 2014; Wurzler et al., 2000). Therefore, developing methods to better integrate aerosol properties into predictive models is essential for improving forecast skill.

In many desert-adjacent regions, mineral dust particles are the most abundant aerosols and serve as the primary source of CCN and Ice Nuclei (IN) (Ganor and Foner, 1996; Levi and Rosenfeld, 1996; Zhang et al., 2021). These particles, often composed of clay minerals like illite and kaolinite, are typically considered poorly hygroscopic (Attwood and Greenslade, 2011; Ganor, 1991). However, they can become effective CCN if coated with water-soluble materials like sulfates during atmospheric transport, or even in their uncoated state under specific thermodynamic conditions (Levin et al., 1996; Wurzler et al., 2000). For these largely insoluble nuclei, the initial activation and condensation of water vapor is a critical rate-limiting step that governs the entire cloud formation process. Understanding this physical mechanism is a prerequisite for accurately modeling cloud development in dust-laden environments.

The characterization of the atmospheric state necessary for such modeling relies on a suite of instruments, each with inherent limitations. Radiosondes provide essential in-situ vertical profiles of temperature, humidity, and wind, but their significant lateral drift means measurements at altitude may be kilometers away from their launch point, reducing their spatial and temporal representativeness for a specific location (Seidel et al., 2011). Conversely, ground-based remote sensing instruments like ceilometers offer continuous, high-resolution monitoring of cloud and aerosol backscatter directly above the instrument, but they lack the thermodynamic context provided by radiosondes (Kotthaus et al., 2023). While these instruments are often used independently, a synergistic approach that combines their respective strengths can unlock new observational and predictive capabilities.

This study introduces and validates a novel framework that helps in predicting cloud formation by synergistically integrating radiosonde and ceilometer data. We present a physical model of condensation on insoluble nuclei and use it to perform a transformation of radiosonde thermodynamic profiles into a simulated optical backscatter signal. By quantitatively comparing this model-derived signal with simultaneously measured ceilometer data, we can determine a key predictive parameter: the minimum effective size of dust particles required to act as CCN under the observed conditions. This work provides a physically validated methodology for using in-situ aerosol data to constrain cloud formation parameterization. By improving the representation of microphysical processes in dust-laden environments, this approach lays the groundwork for more accurate precipitation forecasting - a critical requirement for effective agrometeorological planning and water management.

## 2    Measurements

Atmospheric data was collected by floating ten Meteomodem M10 radiosondes, over a week at the end of November 2017 in the morning, noon, evening and night (Arielly, 2020). The M10 device simultaneously measured lateral location, altitude, horizontal and vertical velocities of the device, Pressure, temperature, relative humidity, and the wind's speed and direction, and transmitted them by radio broadcast to the SR10 receiver device. Measurements were also acquired with Vaisala's CL51 ceilometer system at the same time as the radiosonde data measurements (Arielly, 2020). The ceilometer includes a LASER source that emits $910(10)\ nm$ radiation in $\sim110\ nsec$ pulses. This radiation is back-scattered by the medium in which it moves and absorbed at the detector component which collects 1540 samples at a rate of 15 MHz and produces a two-way attenuated optical backscatter profile ($B$) with spatial sampling resolution of $10\ m$ and a maximum range of $15.4\ km$.

The radiosondes' flight data including the radiosonde number and floating time are shown in **Figure 1** along with horizontal position and elevation (indicated by the color of the trajectory) on top of satellite images (Terrametrics, 2020). While at the beginning of their paths, the radiosondes fly in different horizontal directions, after reaching a height of several kilometers, all the radiosondes flew to the east/northeast. The radiosonde gradually climbs the atmosphere, measuring a series of air parcels at different times, altitudes, and geographical location due to the existence of a finite horizontal wind. Thus, while still in the same lowland geographical area as the starting point, upon reaching a height of ~6 km, the horizontal spread of the radiosondes is ~13 km. This spatial and temporal spread can make the measurements unrepresentative, and is in contrast to the ceilometer's measurement which is acquired almost instantly. Therefore, in order to compare data from these two devices, an equivalent ceilometer optical backscatter profile ($B_m$) was constructed according to each radiosonde flight.

Since the radiosondes all started ascending from roughly the same location where the ceilometer is and are affected from the same wind as the measured air, an air volume measured at a particular geographical point by a radiosonde may have been measured by the ceilometer at a different time. We may use the horizontal wind speed measurements and the ceilometer and radiosondes locations in order to retrace the moments when these air parcels were above the ceilometer. The distance between two locations on the earth's surface is calculated by using the Haversine formula (Sinnott, 1984). The ceilometer's and radiosondes' locations are defined in terms of geographical coordinates as $\vec{r}^{ceil}$ and $\vec{r}^{sonde}(z)$ respectively, along with the following auxiliary function

$$f\big(\vec{r}^{sonde}, \vec{r}^{ceil}\big) = \sin^2\left(\frac{\Delta\vec{r}_y}{2}\right) + \cos(\vec{r}_y^{ceil})\cos(\vec{r}_y^{sonde})\sin^2\left(\frac{\Delta\vec{r}_x}{2}\right) \qquad (1)$$

where $\Delta\vec{r}(z) = \vec{r}^{sonde}(z) - \vec{r}^{ceil}$. Therefor the lateral distance between these two locations is

$$d(z) = 2R_e\tan^{-1}\left(\sqrt{\frac{f(\vec{r}^{sonde}(z), \vec{r}^{ceil})}{1 - f(\vec{r}^{sonde}(z), \vec{r}^{ceil})}}\right) \qquad (2)$$

The travel time will therefore be

$$\Delta t(z) = \frac{d(z)}{V_{Hor}(z)} \qquad (3)$$

where $V_{Hor}(z)$ is the horizontal wind velocity measured by the radiosonde at height $z$, and $R_e$ is Earth's mean radius. For each radiosonde measurement altitude, $z_n$, the data for ranges between $(z_{n-1} + z_n)/2$ and $(z_n + z_{n+1})/2$ from the ceilometer measurement temporally closest to the moment where the air parcel was above it was concatenated to the data corresponding to previous altitudes.

**Figure 2** shows the radiosondes' temperature and relative humidity measurements and the temporally matched ceilometer's measurements ($B_m$). The displayed altitude range is 5 km, which means that the radiosonde data was collected across ~11 km of lateral distance which may complicates this comparison due to uneven conditions and nonlinear trajectories. However, two considerations may resolve this complication: 1) While the optical backscatter data shows very sharp and high peaks which probably originates from the existence of clouds, its signal-to-noise ratio is quite low for heights above 3 km for many backscatter profiles. 2) One can spot subtle changes in temperature that in some panels manifest as local minimum and maximum values accompanied by sharp changes in relative humidity (marked by red arrows. In some panels it is less obvious). These are identified as inversion layers and occur at heights up to ~3 km as well. Considering the altitudes of these inversions reinforced by existing knowledge on the structure of the ABL (Lee et al., 2019; Stull, 1988) together with the altitude dependence of the signal to noise ratio, it is apparent that the most meaningful altitude range for a comparative analysis of data from radiosondes and ceilometer is below 3 km. This means that the flight paths to be considered are all contained within ~4 km of the same lateral geographical area as the starting point, a lowland area, with topographical difference below 100 m.

## 3   Results and discussion

### 3.1   Relating the ceilometer's signal to the scatterers' density

Radiation scattered towards a LiDAR's detector is absorbed, producing an electric signal that is registered as a function of time ($t$) since the production of the LASER pulse. For back-scattered radiation, the range is $z = \frac{1}{2}ct$, and this signal is (Mattis and Wagner, 2014)

$$P(z) = C\ O(z)\alpha(z)S^{-1}(z)/z^2 \exp\left(-2\int_0^z \alpha(z')dz'\right) \qquad (4)$$

where $P(z)$ is the raw detector signal (from which the dark current and atmospheric background signals are subtracted), $O(z)$ is the laser and detector fields of view overlap function, $C$ is a variable that incorporates various characteristic of the system such as the intensity and temporal-width of the laser pulse, the area of the detector, and the quantum efficiency of the detector, $S(z)$ is the LiDAR ratio, and $\alpha(z)$ is the volume extinction coefficient. Each pulse of the ceilometer's LASER provides a vertical two-way attenuated optical backscatter profile, $B(z) = \frac{P(z)z^2}{C\ O(z)}$ (Vaisala, 2017a, 2017b).

The volume extinction coefficient, $\alpha(z) = n_d(z)\sigma_\lambda$, depends on the density of the scatterers, $n_d$, and the radiation extinction cross-section area $\sigma_\lambda$. Assuming spherical scatterers (due to the water droplets surface tension) with radius $r_d$, $\sigma_\lambda = Q_s \pi r_d^2$ is dependent on the geometrical cross section area and on the extinction efficiency, $Q_s$, which like $n_d$ is dependent on $r_d$.

Several attempts have been made to analytically describe measured particle size distributions with simple mathematical expressions (Measures, 1992; Takeuchi, 2005). One of these expressions is the log-normal distribution, in which the size distribution is described by a normal distribution with the particle sizes inserted into the Gaussian in a logarithmic scale. This distribution normalized to the total particle number, $n_{total}$, is

$$n_d(r_d) = \frac{n_{total}}{r_d \sqrt{2\pi}\sigma_d} \exp\left(-\frac{\left(\ln\frac{r_d}{\bar{r}_d}\right)^2}{2\sigma_d^2}\right) \qquad (5)$$

where $r_d$ is the particle radius, $\bar{r}_d$ is the distribution median, and $\sigma_d$ is the distribution width.

Optical scattering from spherical particles with typical size comparable to the wavelength (like some aerosols in the ABL which are sized in the $\sim 0.1 - 10\ \mu m$ range (Mahowald et al., 2014)) is described by the Mie model (Bohren and Huffman, 1983a). This model predicts the probability for scattering the radiation at some angle relative to its original direction, which varies according to the particle size to wavelength ratio, and the optical refractive index of the scattering medium. **Figure 3** shows the dependencies of $n_d$ and $Q_s$ on $r_d$. $Q_s$ (solid line) is calculated from Mie theory for $\lambda = 910\ nm$ and goes to zero for $r_d \to 0$, reaches a maximum value of 4 for $r_d = \lambda$, and converges to 2 for $r_d \gg \lambda$ (Bohren and Huffman, 1983b). Normalized $n_d$ (dashed line) is calculated according to Eq. (5) with $\bar{r}_d = 3.5\ \mu m$ and $\sigma_d = 0.35\ \mu m$ which should fit the particle size distribution in a cumulus cloud which is expected to be the main scatterer, due to the season, lack of precipitations, visual observation, and altitude range (Goldreich, 2003a, 2003b, "Israel Meteorological Service," n.d.; Measures, 1992; Takeuchi, 2005).

Since the ceilometer's wavelength is smaller than this cloud size distribution's typical particle, and under the approximation of mono-dispersity, the cross-section can be set as a constant, $\sigma_\lambda \cong 2\pi\bar{r}_d^2$. For the same reason, $S(z)$ is also a constant (back-scattering from particles identical in size and composition), and $B(z)$ can be written as

$$B(z) \cong A n_d(z) \exp\left(-2\sigma_\lambda \int_0^z n_d(z')dz'\right) \qquad (6)$$

where $A$ is a proportionality constant. Thus, with Eq. (6), it is possible by having a measure of $B$ to have a measure of $n_d$, or more precisely for this work, quantify the mechanism that is responsible for $n_d$.

### 3.2   Formation of scatterers in the atmosphere

As clear from the values of $Q_s$ in **Figure 3**, the ceilometer's measurement is chiefly sensitive to the existence of particles with diameters equal or larger than the ceilometer's wavelength, but the ability of these large particles to reach a certain height decreases with height (Kleinman and Daum, 1991), so due to the absence of molecules other than water that can undergo a phase transition in the discussed heights (Wallace and Hobbs, 2006a, 2006b), the existence of these particles at these heights will result from water adsorption on smaller particles and coalescence of water covered particles (Jonas and Mason, 1974; Kovetz and Olund, 1969), forming a droplet population with $n_d$ density. Since most aerosol particles in the atmosphere over Israel originate from the nearby large desert area and, unless went several stages of atmospheric processing, have poor water adsorption capabilities (Attwood and Greenslade, 2011; Chester, 1990; Cuadros et al., 2015; Falkovich et al., 2001; Ganor, 1991, 1975; Ganor and Foner, 1996; Ganor and Mamane, 1982; Levi and Rosenfeld, 1996; Levin et al., 1990; Wurzler et al., 2000), for a process of water adsorption on these particles to still occur, the water molecules should form a shell around a particle that will hold itself intact with the help of the water molecules' strong hydrogen bonds. This phenomenon is known as a hydrophobic force. It acts when introducing a non-polar solute into water and is caused by the entropy loss minimization to the network of water molecules (Schauperl et al., 2016; Silverstein, 1998; Voet and Voet, 2011). The probability for this process is dependent on the density of gaseous water molecules, $n_v$, which we calculate from the partial pressure of the water vapor in the atmosphere, $P_v$, and the temperature, $T$, by using the ideal gas law

$$n_v = \frac{P_v}{k_B T} \tag{7}$$

where $k_B$ is the Boltzmann constant. The partial pressure of the water vapor in the atmosphere is calculated with (Ahrens and Henson, 2019)

$$P_v = \phi P_{H_2O}^* \tag{8}$$

where $\phi$ is the relative humidity measured by the radiosonde, and $P_{H_2O}^*$ is the water's saturation vapor pressure calculated according to the Buck equation (Buck, 1981; Buck Research Instruments, 2012)

$$P_{H_2O}^*(T) = a \exp\left[\left(b - \frac{T}{d}\right)\left(\frac{T}{c + T}\right)\right] \tag{9}$$

where $a = 6.1121\ mbar$, $b = 18.678$, $c = 257.14\ ^OC$, $d = 234.5\ ^OC$, and $T$ is in degrees Celsius.

Water molecules move through space and scatter off other particles as long as they continue failing to adhere to these particles. As stated above, the water molecules will permanently adhere to a particle if there are enough molecules to form a shell around it, but this has to occur faster than the collision (reorganization) time of the water molecules, $t_c$, and if it succeeds the particle could grow to a size measurable by the ceilometer by absorbing more water vapor or coalescing with other droplets. In a particle system subjected to certain pressure and temperature conditions, $t_c$ can be calculated by again using the ideal gas law together with the known relations $\frac{3}{2}k_B T = \frac{1}{2}mv^2$, $v = l_c/t_c$, $l_c\sigma = 1/n$, $\sigma = \pi r_p^2$, where $v$ is the particles' velocity, $n$ is the particles' density, $m$ is the particles' mass, $l_c$ is the typical distance between the particles, $\sigma$ is the particles' scattering cross section area, and $r_p$ is the particles radius, in order to obtain

$$t_c = \sqrt{\frac{mk_B T}{3P^2\left(\pi r_p^2\right)^2}} \tag{10}$$

For water molecules ($m = m_{H_2O} \cong 18\ amu, 2r_p = 2r_{H_2O} = 2.74$ Å) at room temperature and atmospheric pressure $t_c \cong 1\ nsec$ (D'Arrigo, 1978; Zhang and Xu, 1995).

At the dew temperature, $T_{dew}$, there is an equilibrium between the gaseous and the liquid phase. There will be enough thermal energy to transform from liquid (adsorbed) to gaseous phase only if the excess energy above the thermal energy of the dew temperature is greater than the latent heat energy per particle (the enthalpy), $E_{cond}$, which is

$$E_{cond}(T) = L_W(T)\rho_M/N_A \tag{11}$$

where $\rho_M$ is the molar mass of water, $N_A$ is the Avogadro number, and $L_W$ is the latent heat energy of water, empirically determined as (Rogers and Yau, 1989)

$$L_W(T) = (2500.8 - 2.36T + 0.0016T^2 - 0.00006T^3)\ J/g \tag{12}$$

The dew temperature is calculated with the Magnus equation (Magnus, 1844) which describes the relation between the temperature, the relative humidity, and the water vapor's partial pressure, in its reciprocal form

$$T_{dew} = \frac{c\gamma_m(T, \phi)}{b - \gamma_m(T, \phi)} \tag{13}$$

where $\gamma_m(T, \phi) = \ln(\phi P_{H_2O}^*(T)/a)$, and all other constants are the same as in Eq. (9).

Therefore, considering the Boltzmann factor for this process, for $T \geq T_{dew}$, a water molecule condensation probability is

$$P_{cond} = \exp\left(\frac{k_B(T_{dew} - T)}{E_{cond}}\right) \tag{14}$$

As stated above, in order for the aerosol particles to act as efficient condensation nuclei, an initial water molecules shell (consisting of $\eta$ molecules) should be created around it in a time shorter than $t_c$, a process whose probability is $P_{cond}^\eta$. Considering an abundant source of CCNs of finite sizes (the smallest of which can be fully covered by $N_{mol}$ molecules), the total probability for this process is (assuming $k_B|T_{dew} - T|/E_{cond} \ll 1$)

$$P_T = \sum_{\eta = N_{mol}}^{\infty} P_{cond}^\eta = \frac{-E_{cond}}{k_B(T_{dew} - T)}\exp\left(N_{mol}\frac{k_B(T_{dew} - T)}{E_{cond}}\right) \tag{15}$$

Therefore, the drop density is

$$n_d = n_v P_T \tag{16}$$

### 3.3   Model Validation and Parameter Estimation

Using the above relations, the radiosondes measurements were transformed into simulated optical back-scattered curves and fitted to the ceilometer's measurement by adjusting $N_{mol}$. **Figure 4** shows a comparison of $n_d$ calculated using Eq. (16), $B'$ calculated using Eq. (6), and $B_m$ measured by the ceilometer and temporally adjusted for the radiosonde measurements as explained above.

There seems to be a high visual correlation between $B_m$ and $B'$, which is mostly well confirmed by calculated correlation values, $\rho_{B_m, B'}$, between the two (**Table 1**). This confirms the physical basis for the model presented here. This strong agreement is a critical first step, as it demonstrates that the model is reliable enough to serve as the foundation for a predictive tool. The model also allows us to estimate the minimum condensation nuclei size of the droplets from $N_{mol}$ by considering the effective coverage area of "spherical" water molecules on "spherical" condensation nuclei

$$r_{nuc} = \left( \frac{N_{mol} \pi r_{H_2O}^2}{4 \pi f_{pack}} \right)^{1/2} = \sqrt{\frac{N_{mol}}{4 f_{pack}}} r_{H_2O} \qquad (17)$$

where $r_{H_2O}$ is the water molecule radius and $f_{pack} \cong 0.9069$ is the molecules' packing ratio on top of the particle's surface (Conway and Sloane, 1999). Values for $r_{nuc}$ from these comparisons are shown in **Table 1**, and have a mean value of $2r_{nuc} = 46.7(18) \ nm$. This size does indeed correspond to the known minimum size of cloud condensation nuclei (Mahowald et al., 2014). This small size seems counter intuitive since, due to the curvature, it is expected that the existence of smaller droplets will require higher super-saturation values to maintain thermodynamic equilibrium, according to the Kelvin equation (Seinfeld and Pandis, 2016a). For soluble CCN, this is not necessarily true, and an opposite dependence between the droplet size and super-saturation is expected for the smallest CCN sizes according to Köhler theory (Köhler, 1936; Seinfeld and Pandis, 2016b). In our case, however, the CCNs are assumed to be non-soluble and non-evaporate (Attwood and Greenslade, 2011; Falkovich et al., 2001; Ganor, 1991), and the kelvin equation and Köhler theory are not applicable for the creation of the first water layer, which we assume is the rate determining step. The effect of the estimated CCN size curvature on the condensation probability due to surface tension (Berry, 1971) was estimated to be $\Delta P_{cond}/P_{cond} < 0.007\%$.

The most sensitive part in estimating the $r_{nuc}$ values is the transformation of the radiosonde measurements into simulated backscatter values. These measurements have their uncertainties which may affect the calculated values. Considering the ranges of measured values and the accuracies stated by the manufacturer (Meteomodem, n.d.), the leading uncertainty causing element is the relative humidity measurement with 4-6% relative uncertainty at the considered altitudes. This is actually below the standard deviation value of $r_{nuc}$ of ~12%. While a 12% standard deviation in the derived $r_{nuc}$ is acceptable for this validation study, understanding its impact is crucial for operational use. For example, a key question for future work is determining the threshold at which this level of uncertainty would alter a 'rain/no-rain' forecast. However, the fact that the primary source of uncertainty (RH sensor) does not create runaway errors in the output ($r_{nuc}$) demonstrates the model's fundamental stability for applied forecasting.

Furthermore, although one would expect the ceilometer's signal to always correspond to the water droplet density, it often displays a strong signal where the density is relatively low, and a weak signal where the density is high. This, of course, is due to the exponent term in Eq. (4) which more strongly attenuates the signal for greater distances.

While it seems that this model successfully describes most of the ceilometer's measurements features, some measurements contain a feature that is not described by the model, and therefore their quality of fitting is lower (for example – panel 1 in Figure 4). This feature is pronounced in measurements with relatively low signal to noise ratios and is located at heights of below 1 km.

**Table 1.** Correlation values, $\rho_{B_m,B'}$, and condensation nuclei diameters, $2r_{nuc}$, calculated for each of the radiosondes. $\rho_{B_m,B'}$ values are usually high, except for measurements where the signal to noise ratio is low, and their average value is 0.85(5). $2r_{nuc}$ values are narrowly distributed around an average value of 46.7(18) nm corresponding to the known minimum size of cloud condensation nuclei.

| Radiosonde number | $\rho_{B_m,B'}$ | $2r_{nuc}$ |
|---|---|---|
| 0 | 0.99 | 44.1 |
| 1 | 0.65 | 48.8 |
| 2 | 0.75 | 42.4 |
| 3 | 0.81 | 56.6 |
| 4 | 0.97 | 37.5 |
| 5 | 0.92 | 43.9 |
| 6 | 0.52 | 47.7 |
| 7 | 0.99 | 51.9 |
| 8 | 0.97 | 51.1 |
| 9 | 0.90 | 43.2 |

The feature is shaped like an asymmetrical bell curve centered around 0.5 km, which can correspond to finite wavelength-sized aerosol concentrations expected at low altitudes (Kleinman and Daum, 1991), together with non-constant LiDAR overlap function (Eq. (4)). Another explanation for these anomalies could be scattered LASER radiation contribution of second or higher order that is becoming significant for sensing ranges that are equal or smaller than the visibility distance or than the source-detector distance of the ceilometer (Ding et al., 2010). These low-altitude anomalies help define the potential operational envelope of the model. They indicate that for robust predictions at low altitudes (<1 km), the model may need to be augmented. For cloud formation above this level, the model remains valid and accurate. This clarifies the conditions under which this framework can be reliably applied for forecasting.

### 3.4   Implications for Applied Forecasting and Water Management

While the preceding analysis focused on validating our model against atmospheric data, the following discussion serves a broader purpose: to establish the model's credibility for predictive applications. Having confirmed that the model accurately captures the key physics of droplet nucleation on dust, we discuss its potential implementation in operational meteorology. Translation of CCN characteristics into actionable precipitation forecasts would allow agricultural stakeholders to potentially better plan irrigation schedules and optimize resource allocation based on the likelihood of natural rainfall events.

To leverage this model effectively for rainfall probability estimates, certain adjustments and parameter assessments are necessary. We can use Eq. (17) in order to describe Eq. (15) as

$$P_T(T,\phi) = \frac{1}{g(T,\phi)} \exp\left(-4f_{pack}\left(\frac{r_{nuc}}{r_{H_2O}}\right)^2 g(T,\phi)\right) \qquad (18)$$

where $g(T, \phi) = -\frac{k_B(T_{dew} - T)}{E_{cond}}$. Due to the minus sign in the exponent and the positive values of $g$, this probability expression will contribute the most to rainfall when $g$ and $r_{nuc}$ are as small as possible. Investigation of $g$ values in our measured dataset concludes that up to heights of 2 km, the minimum value of $g$ will be $3.2(3) \times 10^{-4}$, i.e. the mean error is less than 10%. Thus, it is possible to take $g$ as a constant in the model and only calibrate it between seasons to account for the variations in atmospheric thermodynamics. Consequently, the primary variable that would require continuous monitoring is the CCN size distribution, particularly identifying the smallest effective CCN size, $r_{nuc}$, for reliable droplet nucleation. Operational forecast models would be able to use the measured minimal CCN size as input for the seasonal calibrated model in order to calculate rain event probability.

This method's primary advantage lies in its simplicity and minimal data requirements, as it relies on a single, regularly measured variable - the CCN size spectrum. By using a particle size analyzer, accurate measurements of CCN sizes can be obtained in real time, focusing on the smallest effective CCN size required for droplet nucleation. This straightforward approach enables local and timely assessments. While condensation is a prerequisite and not a guarantee of rainfall, once local measurements indicate that dust particles are of an effective size for condensation, the possible subsequent process of droplet growth into raindrops is known to take a minimum of 25 minutes in an aerosol-rich environments (Jonas and Mason, 1974). This inherent physical delay provides a crucial and concrete lead time for operational decision-making, making the model advantageous for on-the-ground applications. By streamlining the model's inputs in this way, the approach becomes beneficial for rain event prediction, aligning with sustainable water management strategies that are particularly valuable in agriculture-dependent, desert-adjacent regions.

## 4    Conclusions

This study demonstrates that the initial formation of water shells on insoluble mineral dust - a critical rate limiting step in arid environments, can be successfully modeled and validated using a synergy of radiosonde and ceilometer data. By transforming thermodynamic profiles into simulated optical backscatter and comparing them with ceilometer data, we identified the minimum effective CCN size required for activation. The agreement between the model and observed ceilometer signals confirms that insoluble dust particles play a predictable and quantifiable role in local cloud formation.

The implications of this validation extend beyond microphysical theory to operational meteorology. In arid and semi-arid regions, where dust is the dominant aerosol, the ability to constrain the specific conditions for droplet nucleation offers a pathway to reduce the uncertainty in precipitation forecasting. While this study focused on the physical validation of the condensation mechanism, the framework provides the necessary physical basis for developing real-time, aerosol-aware forecast tools. Future work should include testing this method using monitored rain events backed by continuous particle size measurements, and future applications could include integration into automated systems for real-time precipitation forecasting, enhancing the ability of climate-sensitive sectors, such as agriculture, to make data-driven decisions amid shifting environmental conditions.

**Conflicts of interests and Data**

The authors declare no financial conflicts of interests. The data supporting the conclusions of this work can be obtained at the Mendeley Data server (Arielly, 2020).

**Funding**

This research did not receive any specific grant from funding agencies in the public, commercial, or not-for-profit sectors.

**Author Contributions**

Rani Arielly: Conceptualization, Data curation, Formal analysis, Investigation, Methodology, Visualization, Writing – original draft;

Adva Baratz: Conceptualization, Writing – review and editing;

Ran Aharoni: Validation, Writing – review and editing;

Ofir Shoshanim: Conceptualization, Writing – review and editing

**Declaration of generative AI and AI-assisted technologies in the manuscript preparation process**

During the preparation of this work the author(s) used Gemini and NotebookLM in order to refine the text and improve its readability. After using this tool/service, the author(s) reviewed and edited the content as needed and take(s) full responsibility for the content of the published article.

# References:

Ahrens, C.D., Henson, R., 2019. Humidity, in: Meteorology Today: An Introduction to Weather, Climate, and the Environment. Cengage Learning, Boston, MA, USA, pp. 95–107.

Arielly, R., 2020. Ceilometer and Radiosonde Data - November 2017 [dataset]. https://doi.org/10.17632/xrpshxvt8w.1

Attwood, A.R., Greenslade, M.E., 2011. Optical Properties and Associated Hygroscopicity of Clay Aerosols. Aerosol Sci. Technol. 45, 1350–1359. https://doi.org/10.1080/02786826.2011.594462

Berry, M.V., 1971. The molecular mechanism of surface tension. Phys. Educ. 6, 79–84. https://doi.org/10.1088/0031-9120/6/2/001

Bohren, C.F., Huffman, D.R., 1983a. Absorption and Scattering by a Sphere, in: Absorption and Scattering of Light by Small Particles. John Wiley & Sons, Inc., New York, pp. 82–129.

Bohren, C.F., Huffman, D.R., 1983b. The Extinction Paradox; Scalar Diffraction Theory, in: Absorption and Scattering of Light by Small Particles. John Wiley & Sons, Inc., New York, pp. 107–111.

Buck, A.L., 1981. New equations for computing vapor pressure and enhancement factor. J. Appl. Meteorol. 20, 1527–1532.

Buck Research Instruments, L., 2012. Model CR-1A Hygrometer With Autofill Operating Manual.

Cecchini, M.A., Machado, L.A.T., Artaxo, P., 2014. Droplet Size Distributions as a function of rainy system type and Cloud Condensation Nuclei concentrations. Atmospheric Res. 143, 301–312. https://doi.org/10.1016/j.atmosres.2014.02.022

Chester, R., 1990. The atmospheric transport of clay minerals to the world ocean, in: Proceedings of the 9th International Clay Conference. Strasbourg, pp. 23–32.

Conway, J.H., Sloane, N.J.A., 1999. The Sphere Packing Problem, in: Berger, M., Coates, J., Varadhan, S.R.S. (Eds.), Sphere Packings, Lattices and Groups. Springer, New York, pp. 1–21.

Cuadros, J., Diaz-Hernandez, J.L., Sanchez-Navas, A., Garcia-Casco, A., 2015. Role of clay minerals in the formation of atmospheric aggregates of Saharan dust. Atmos. Environ. 120, 160–172. https://doi.org/10.1016/j.atmosenv.2015.08.077

D'Arrigo, J.S., 1978. Screening of membrane surface charges by divalent cations: an atomic representation. Am. J. Physiol.-Cell Physiol. 235, C109–C117. https://doi.org/10.1152/ajpcell.1978.235.3.C109

Ding, H., Xu, Z., Sadler, B.M., 2010. A Path Loss Model for Non-Line-of-Sight Ultraviolet Multiple Scattering Channels. EURASIP J. Wirel. Commun. Netw. 2010, 598572. https://doi.org/10.1155/2010/598572

Falkovich, A.H., Ganor, E., Levin, Z., Formenti, P., Rudich, Y., 2001. Chemical and mineralogical analysis of individual mineral dust particles. J. Geophys. Res. Atmospheres 106, 18029–18036. https://doi.org/10.1029/2000JD900430

Ganor, E., 1991. The composition of clay minerals transported to Israel as indicators of Saharan dust emission. Atmospheric Environ. Part Gen. Top. 25, 2657–2664. https://doi.org/10.1016/0960-1686(91)90195-D

Ganor, E., 1975. Atmospheric Dust in Israel — Sedimentological and Meteorological Analysis of Dust Deposition (Phd). Hebrew University, Jerusalem, Israel.

Ganor, E., Foner, H.A., 1996. The Mineralogical and Chemical Properties and the Behaviour of Aeolian Saharan Dust Over Israel, in: Guerzoni, S., Chester, R. (Eds.), The Impact of Desert Dust Across the Mediterranean, Environmental Science and Technology Library. Springer Netherlands, Dordrecht, pp. 163–172. https://doi.org/10.1007/978-94-017-3354-0_15

Ganor, E., Mamane, Y., 1982. Transport of Saharan dust across the eastern Mediterranean. Atmospheric Environ. 1967 16, 581–587. https://doi.org/10.1016/0004-6981(82)90167-6

Goldreich, Y., 2003a. Precipitation, in: The Climate Of Israel - Observations, Research and Application. Springer, New York, pp. 55–91.

Goldreich, Y., 2003b. Spatial Cloudiness And Temporal Distribution, in: The Climate Of Israel - Observations, Research and Application. Springer, New York, pp. 127–129.

Israel Meteorological Service [WWW Document], n.d. URL https://ims.data.gov.il/ (accessed 7.10.20).

Jiang, M., Feng, J., Li, Z., Sun, R., Hou, Y.-T., Zhu, Y., Wan, B., Guo, J., Cribb, M., 2017. Potential influences of neglecting aerosol effects on the NCEP GFS precipitation forecast. Atmospheric Chem. Phys. 17, 13967–13982. https://doi.org/10.5194/acp-17-13967-2017

Jonas, P.R., Mason, B.J., 1974. The evolution of droplet spectra by condensation and coalescence in cumulus clouds. Q. J. R. Meteorol. Soc. 100, 286–295. https://doi.org/10.1002/qj.49710042503

Kleinman, L.I., Daum, P.H., 1991. Vertical distribution of aerosol particles, water vapor, and insoluble trace gases in convectively mixed air. J. Geophys. Res. Atmospheres 96, 991–1005. https://doi.org/10.1029/90JD02117

Köhler, H., 1936. The nucleus in and the growth of hygroscopic droplets. Trans. Faraday Soc. 32, 1152–1161. https://doi.org/10.1039/TF9363201152

Kotthaus, S., Bravo-Aranda, J.A., Collaud Coen, M., Guerrero-Rascado, J.L., Costa, M.J., Cimini, D., O'Connor, E.J., Hervo, M., Alados-Arboledas, L., Jiménez-Portaz, M., Mona, L., Ruffieux, D., Illingworth, A., Haeffelin, M., 2023. Atmospheric boundary layer height from ground-based remote sensing: a review of capabilities and limitations. Atmospheric Meas. Tech. 16, 433–479. https://doi.org/10.5194/amt-16-433-2023

Kovetz, A., Olund, B., 1969. The Effect of Coalescence and Condensation on Rain Formation in a Cloud of Finite Vertical Extent. J. Atmospheric Sci. 26, 1060–1065. https://doi.org/10.1175/1520-0469(1969)026%253C1060:TEOCAC%253E2.0.CO;2

Lee, J., Hong, J.W., Lee, K., Hong, J., Velasco, E., Lim, Y.J., Lee, J.B., Nam, K., Park, J., 2019. Ceilometer Monitoring of Boundary-Layer Height and Its Application in Evaluating the Dilution Effect on Air Pollution. Bound.-Layer Meteorol. https://doi.org/10.1007/s10546-019-00452-5

Levi, Y., Rosenfeld, D., 1996. Ice Nuclei, Rainwater Chemical Composition, and Static Cloud Seeding Effects in Israel. J. Appl. Meteorol. 1988-2005 35, 1494–1501.

Levin, Z., Ganor, E., Gladstein, V., 1996. The Effects of Desert Particles Coated with Sulfate on Rain Formation in the Eastern Mediterranean. J. Appl. Meteorol. 35, 1511–1523. https://doi.org/10.1175/1520-0450(1996)035%253C1511:TEODPC%253E2.0.CO;2

Levin, Z., Price, C., Ganor, E., 1990. The contribution of sulfate and desert aerosols to the acidification of clouds and rain in Israel. Atmospheric Environ. Part Gen. Top. 24, 1143–1151. https://doi.org/10.1016/0960-1686(90)90079-3

Magnus, G., 1844. Versuche über die Spannkräfte des Wasserdampfs. Ann. Phys. 137, 225–247. https://doi.org/10.1002/andp.18441370202

Mahowald, N., Albani, S., Kok, J.F., Engelstaeder, S., Scanza, R., Ward, D.S., Flanner, M.G., 2014. The size distribution of desert dust aerosols and its impact on the Earth system. Aeolian Res. 15, 53–71. https://doi.org/10.1016/j.aeolia.2013.09.002

Mattis, I., Wagner, F., 2014. E-PROFILE: Glossary of lidar and ceilometer variables. Department of Research and Development, Meteorological Observatory Hohenpeißenberg, German Weather Service (DWD), Hohenpeißenberg, Germany.

Measures, R.M., 1992. Volume Scattering Coefficients and Phase Functions, in: Laser Remote Sensing: Fundamentals and Applications. Krieger publishing company, Malabar, Florida, pp. 53–58.

Meteomodem, n.d. M10 Radiosonde.

Rogers, R.R., Yau, M.K., 1989. Clausius-Clapeyron equation, in: Ter-Haar, D. (Ed.), A Short Course in Cloud Physics. Butterworth-Heinemann, Burlington, Massachusetts, pp. 12–16.

Schauperl, M., Podewitz, M., Waldner, B.J., Liedl, K.R., 2016. Enthalpic and Entropic Contributions to Hydrophobicity. J. Chem. Theory Comput. 12, 4600–4610. https://doi.org/10.1021/acs.jctc.6b00422

Seidel, D.J., Sun, B., Pettey, M., Reale, A., 2011. Global radiosonde balloon drift statistics. J. Geophys. Res. 116, D07102. https://doi.org/10.1029/2010JD014891

Seinfeld, J.H., Pandis, S.N., 2016a. Equilibrium Vapor Pressure Over A Curved Surface: The Kelvin Effect, in: Atmospheric Chemistry and Physics: From Air Pollution to Climate Change. Wiley & Sons, New Jersey, pp. 419–423.

Seinfeld, J.H., Pandis, S.N., 2016b. Equilibrium Of Water Droplets In The Atmosphere, in: Atmospheric Chemistry and Physics: From Air Pollution to Climate Change. Wiley & Sons, New Jersey, pp. 708–719.

Silverstein, T.P., 1998. The Real Reason Why Oil and Water Don't Mix. J. Chem. Educ. 75, 116. https://doi.org/10.1021/ed075p116

Sinnott, R.W., 1984. Virtues of the Haversine. Sky Telesc. 68, 159.

Stull, R.B., 1988. Mean Boundary Layer Characteristics, in: An Introduction to Boundary Layer Meteorology. Kluwer Academic Publishers, Dordrecht, pp. 2–19.

Takeuchi, N., 2005. Model of Aerosol Size Distribution, in: Fujii, T., Fukuchi, T. (Eds.), Laser Remote Sensing. CRC press, Boca Raton, Florida, pp. 92–95.

Terrametrics, 2020. Landsat-7 image courtesy of the U.S. Geological Survey.

Vaisala, 2017a. Introduction to Vaisala Ceilometer CL51, in: Vaisala Ceilometer CL51 User Guide. Vaisala Oyj, Vantaa, pp. 19–20.

Vaisala, 2017b. Data Messages, in: Vaisala Ceilometer CL51 User Guide. Vaisala Oyj, Vantaa, pp. 60–77.

Voet, D., Voet, J.G., 2011. Protein Stability, in: Biochemistry. Wiley & Sons, NJ, pp. 259–266.

Wallace, J.M., Hobbs, P.V., 2006a. Chemical Composition, in: DMOWSKA, R., HARTMANN, D., ROSSBY, H.T. (Eds.), Atmospheric Science - An Introductory Survey. Academic Press, London, pp. 8–9.

Wallace, J.M., Hobbs, P.V., 2006b. Composition of Tropospheric Air, in: DMOWSKA, R., HARTMANN, D., ROSSBY, H.T. (Eds.), Atmospheric Science - An Introductory Survey. Academic Press, London, pp. 153–157.

Wurzler, S., Reisin, T.G., Levin, Z., 2000. Modification of mineral dust particles by cloud processing and subsequent effects on drop size distributions. J. Geophys. Res. Atmospheres 105, 4501–4512. https://doi.org/10.1029/1999JD900980

Zhang, Y., Xu, Z., 1995. Atomic radii of noble gas elements in condensed phases. Am. Mineral. 80, 670–675. https://doi.org/10.2138/am-1995-7-803

Zhang, Y., Yu, F., Luo, G., Fan, J., Liu, S., 2021. Impacts of long-range-transported mineral dust on summertime convective cloud and precipitation: a case study over the Taiwan region. Atmospheric Chem. Phys. 21, 17433–17451. https://doi.org/10.5194/acp-21-17433-2021

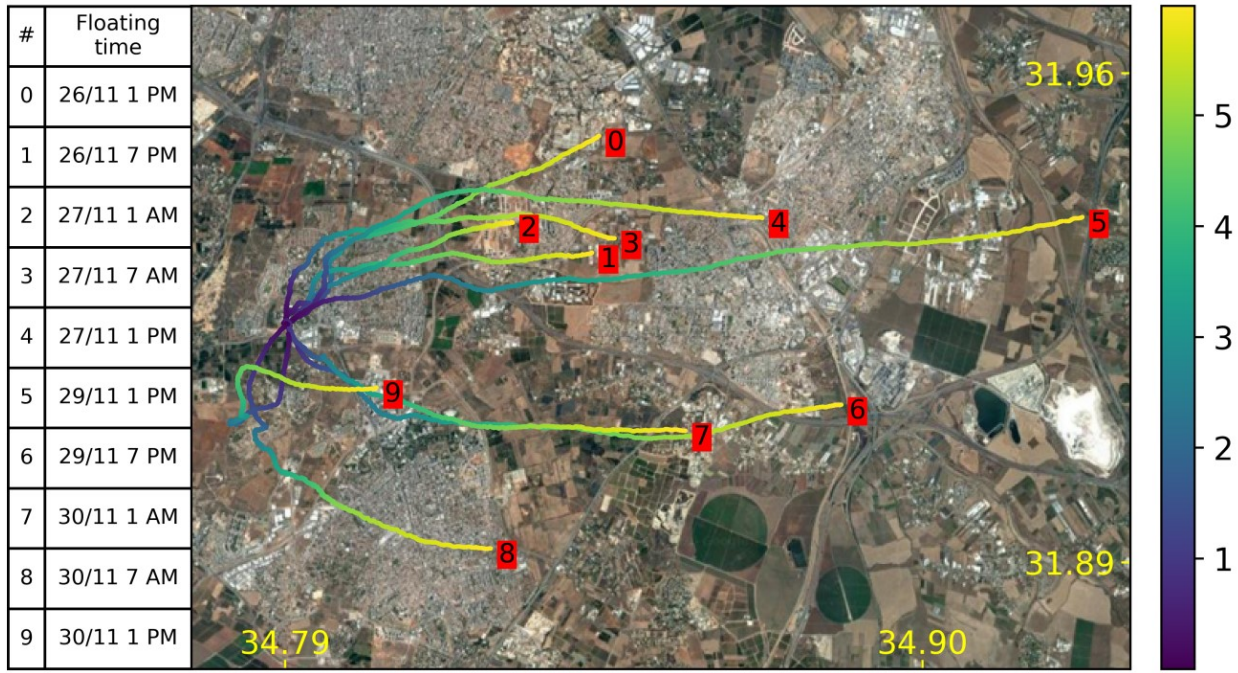

| # | Floating time |
|---|---|
| 0 | 26/11 1 PM |
| 1 | 26/11 7 PM |
| 2 | 27/11 1 AM |
| 3 | 27/11 7 AM |
| 4 | 27/11 1 PM |
| 5 | 29/11 1 PM |
| 6 | 29/11 7 PM |
| 7 | 30/11 1 AM |
| 8 | 30/11 7 AM |
| 9 | 30/11 1 PM |

**Figure 1**. Flight paths of the radiosondes up to altitudes of ~6 km sketched over a space of geographical coordinates of longitude and latitude with a satellite base image. The different trajectories are labeled by numbers in accordance with the table on the left which list their floating times. The color of the line describing a trajectory varies according to the height of the radiosonde at the same location and is explained by the color bar on the right (in km). map lines delineate study areas and do not necessarily depict accepted national boundaries.

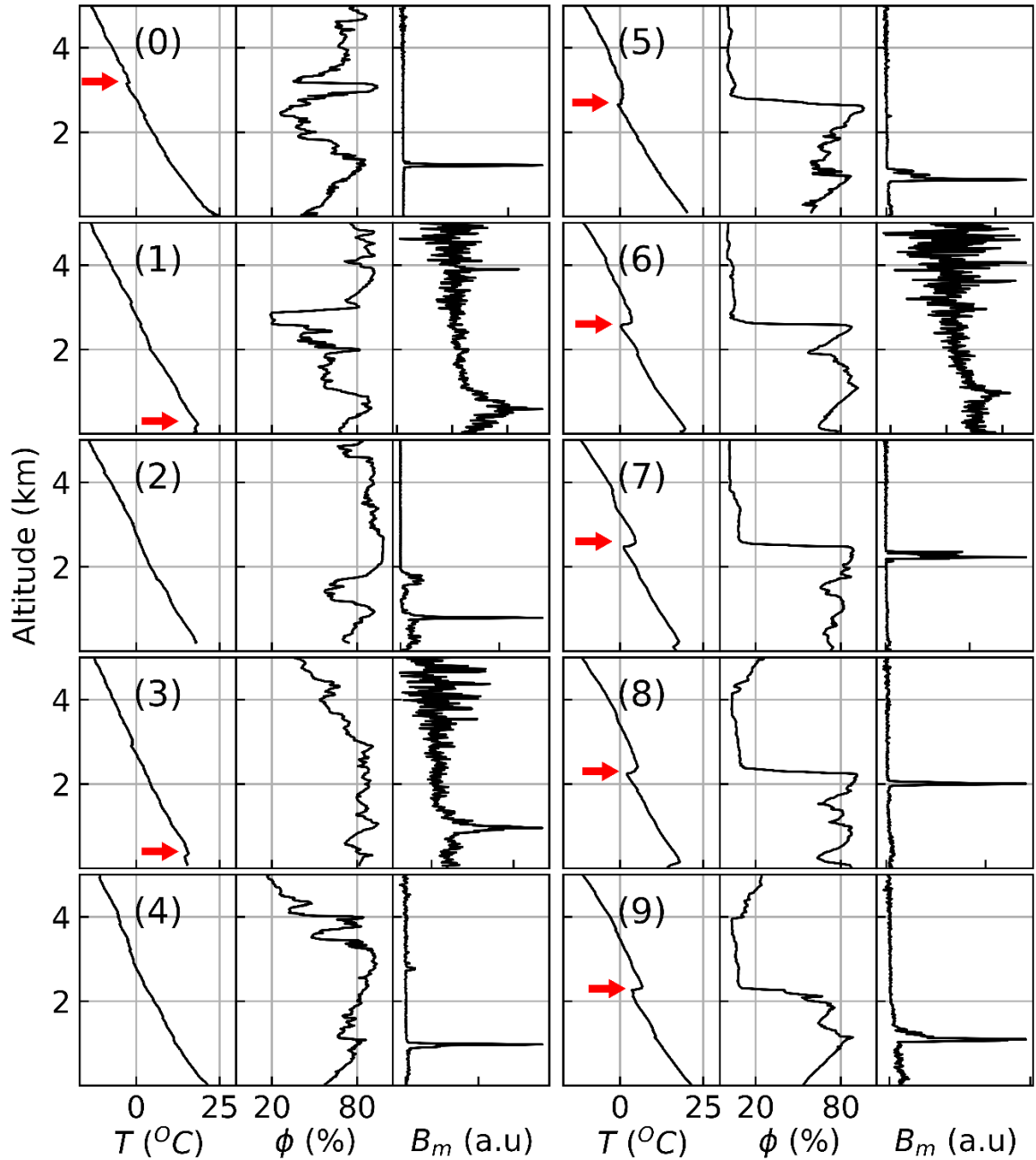

**Figure 2**. Temperature ($T$) and relative humidity ($\phi$) data measured by each of the radiosondes vs. the temporally matched ceilometer measurements ($B_m$). The figure is divided into ten panels, each showing data from a single radiosonde, numbered according to the radiosondes numbers as in **Figure 1**, and divided into three sub-panels for temperature, relative humidity and optical back-scatter measurements.

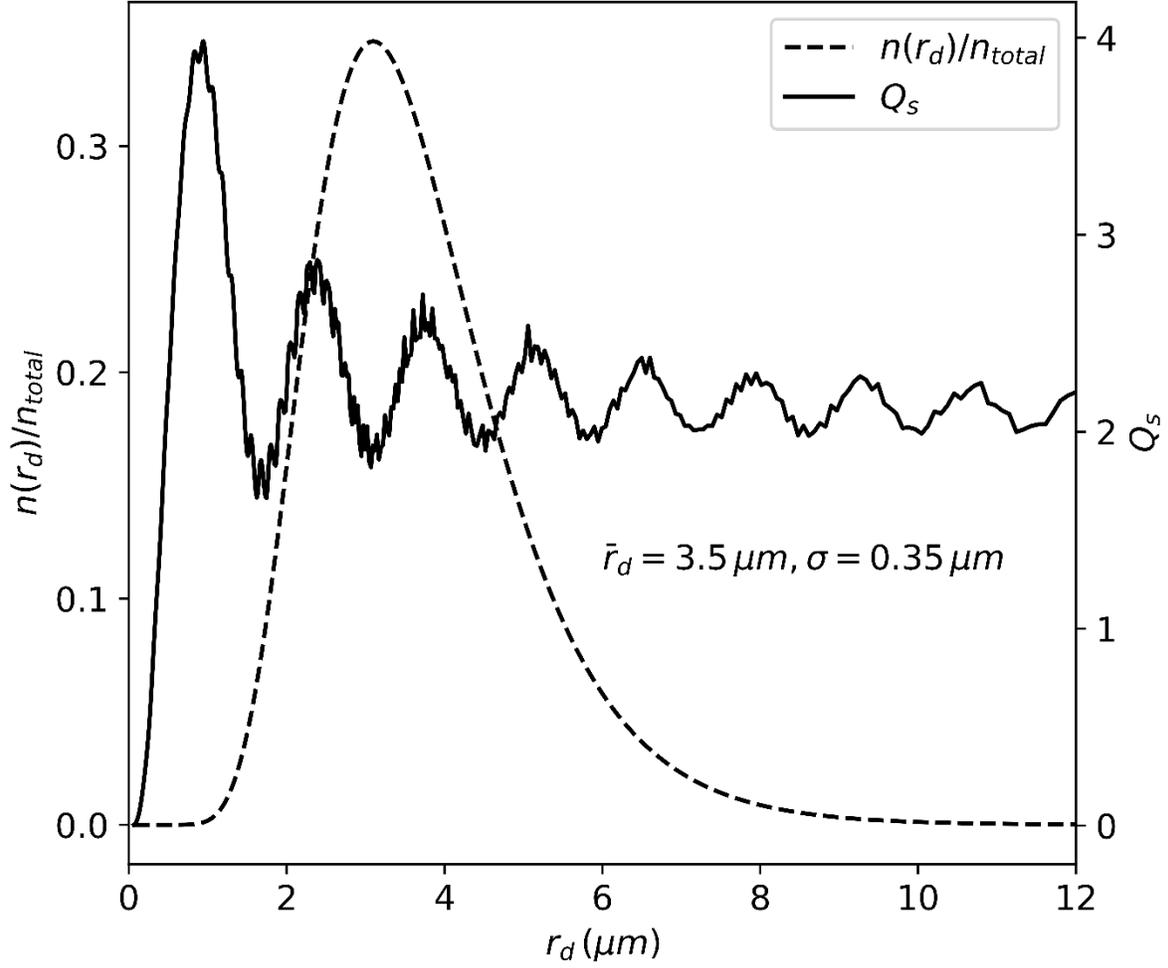

**Figure 3**. Normalized distribution of cloud droplet sizes, $n(r_d)/n_{total}$, (dashed line) and dependence of radiation extinction efficiency, $Q_s$, at wavelength of $\lambda = 910\ nm$ (continuous line) in droplet size, $r_d$. The droplet sizes are log-normal distributed, with the distribution's median at 3.5 μm and its width is 0.35 μm.

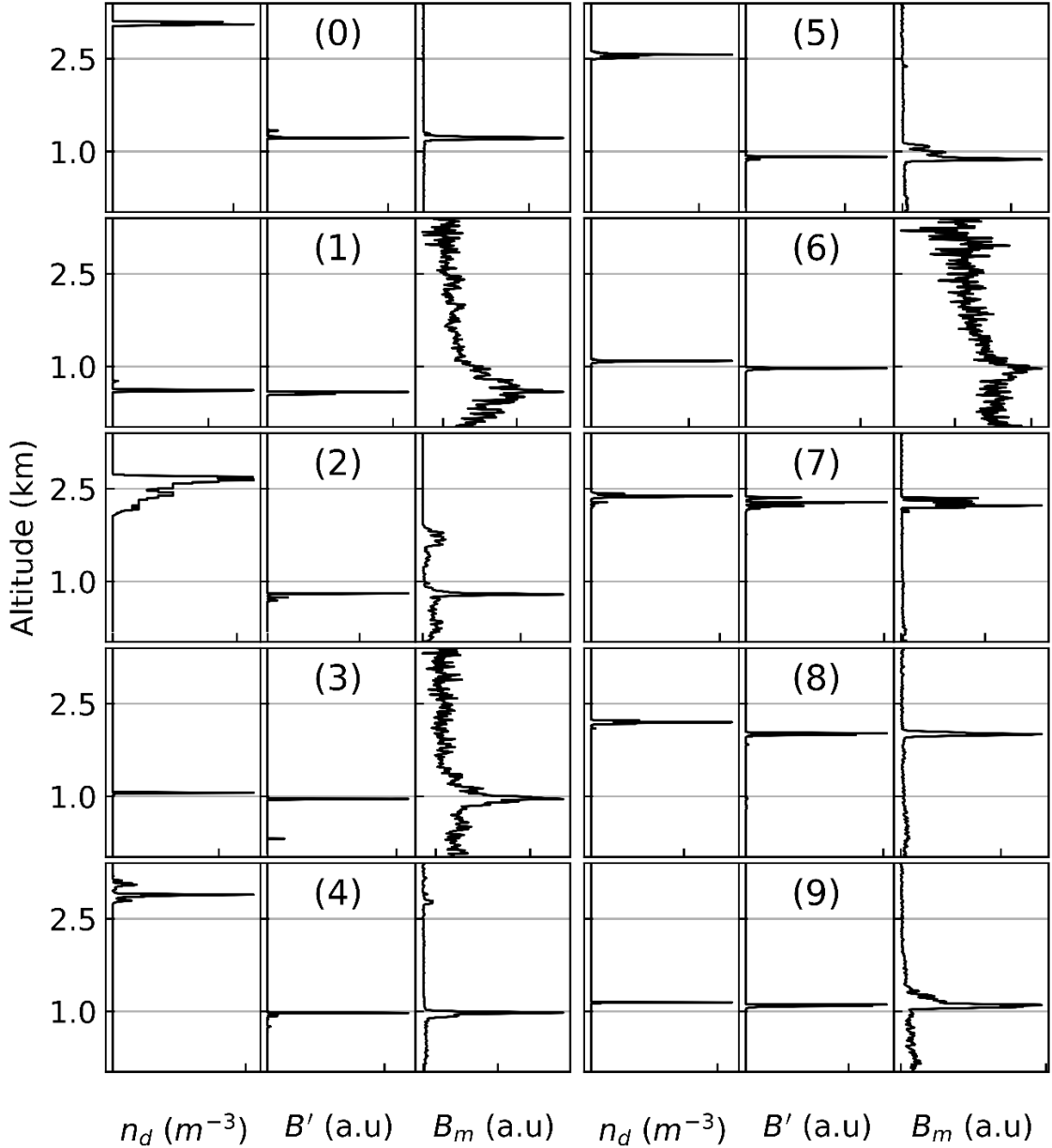

**Figure 4**. Comparison between the calculated drop densities, $n_d$, simulated ceilometer profiles, $B'$, and measured temporally matched ceilometer profiles, $B_m$. The altitude range was set to 3.4 km according to the identified inversion altitudes and to the signal to noise ratio (see section 2). The figure is divided into ten panels, each showing data corresponding to a single radiosonde, numbered according to the radiosondes numbers as in **Figure 1**, and divided into three sub-panels for $n_d$, $B'$ and $B_m$. There is usually a good fit between $B'$ and $B_m$, so that even small features are successfully reproduced, such as secondary peaks near strong peaks in the signal. Correlation based fit values are shown in **Table 1**. When the signal-to-noise ratio is low, there is a significant contribution at low altitudes that is not reproduced by the model and has several explanations (see section 3.3). In four measurements (40%) the sharp peaks' altitudes in the ceilometer signal do not correspond to the altitudes where the droplet density is highest, due to finite droplet density values which are relatively amplified due to the exponent term in the LiDAR equation.